\documentclass[]{spie}  

\usepackage{amsmath,amsfonts,amssymb}
\usepackage{graphicx}
\usepackage[colorlinks=true, allcolors=blue]{hyperref}
\usepackage{float}
\usepackage{booktabs}
\usepackage{wrapfig}
\usepackage{subcaption}

\title{Design of the vacuum high contrast imaging testbed for CDEEP, the Coronagraphic Debris and Exoplanet Exploring Pioneer}

\author[a]{Erin R. Maier}
\author[a]{Ewan S. Douglas}
\author[a,b,c]{Dae Wook Kim}
\author[a]{Kate Su}
\author[b]{Jaren N. Ashcraft}
\author[a,d]{James B. Breckinridge}
\author[b,c]{Heejoo Choi}
\author[e]{Elodie Choquet}
\author[a]{Thomas E. Connors}
\author[a]{Olivier Durney}
\author[a]{Kerry L. Gonzales}
\author[b]{Charlotte E. Guthery}
\author[f]{Christian A. Haughwout}
\author[b]{James C. Heath}
\author[a]{Justin Hyatt}
\author[b]{Jennifer Lumbres}
\author[a]{Jared R. Males}
\author[g]{Elisabeth C. Matthews}
\author[b]{Kian Milani}
\author[a]{Oscar M. Montoya}
\author[h]{Mamadou N'Diaye}
\author[a]{Jamison Noenickx}
\author[f]{Leonid Pogorelyuk}
\author[i]{Garreth Ruane} 
\author[a]{Glenn Schneider}
\author[b]{George A. Smith}
\author[j]{Christopher C. Stark}

\affil[a]{Department of Astronomy and Steward Observatory, University of Arizona, 933 N. Cherry Ave., Tucson, AZ 85719, USA}
\affil[b]{James C. Wyant College of Optical Sciences, University of Arizona, Meinel Building 1630 E. University Blvd., Tucson, AZ. 85721}
\affil[c]{Large Binocular Telescope Observatory, University Of Arizona, 933 N. Cherry Ave. Tucson, AZ 85721}
\affil[d]{Department of Astronomy, California Institute of Technology, 1216 E. California Blvd., Pasadena, CA 91125}
\affil[e]{Aix Marseille Universit\'{e}, CNRS, CNES, LAM, Marseille, France}
\affil[f]{STAR Lab, Department of Aeronautics and Astronautics, Massachusetts Institute of Technology, Cambridge, MA 02139}
\affil[g]{Observatoire de l’Universit\'{e} de Gen\`{e}ve, Chemin Pegasi 51, 1290 Versoix, Switzerland}
\affil[h]{Universit\'{e} C\^{o}te d'Azur, Observatoire de la C\^{o}te d’Azur, CNRS, Laboratoire Lagrange, Nice, France}
\affil[i]{Jet Propulsion Laboratory, California Institute of Technology, Pasadena, CA 91109}
\affil[j]{NASA Goddard Space Flight Center, Greenbelt, MD 20771}

\authorinfo{Send correspondence to Erin R. Maier. E-mail: erinrmaier@email.arizona.edu}

\begin{document} 
\maketitle

\begin{abstract}
The Coronagraphic Debris Exoplanet Exploring Payload (CDEEP) is a Small-Sat mission concept for high contrast imaging of circumstellar disks. CDEEP is designed to observe disks in scattered light at visible wavelengths at a raw contrast level of $10^{-7}$ per resolution element ($10^{-8}$ with post processing). This exceptional sensitivity will allow the imaging of transport dominated debris disks, quantifying the albedo, composition, and morphology of these low-surface brightness disks. CDEEP combines an off-axis telescope, microelectromechanical systems (MEMS) deformable mirror, and a vector vortex coronagraph (VVC). This system will require rigorous testing and characterization in a space environment. We report on the CDEEP mission concept, and the status of the vacuum-compatible CDEEP prototype testbed currently under development at the University of Arizona, including design development and the results of simulations to estimate performance. 
  
\end{abstract}

\newpage

\section{INTRODUCTION}

Over the past quarter century, we have progressed from possessing evidence for only one planetary system, our own, to thousands confirmed today\cite{winn_occurrence_2015}. Studying debris disks at large separations offers an alternative method to characterize planetary system architectures and their evolution which is complementary to the techniques of exoplanet detection \cite{winn_occurrence_2015,wyatt_evolution_2008,krivov_debris_2010,matthews_observations_2014,hughes_debris_2018}.

These disks, identified around stars through their excess of infrared (IR) emission, are composed of dust grains ranging from $\sim$ $\mu$m to mm-size which are continually replenished by sublimation and collisions of planetesimals leftover from planet formation. Sensitive IR surveys provided by space telescopes (e.g., \emph{IRAS}, \emph{Spitzer}, \emph{Herschel}, and \emph{WISE}) have identified thousands of IR excesses around mature stars. Most of these debris disks are not spatially resolved and so their global properties are characterized by the ratio of the integrated dust excess to the luminosity of the star, $L_{IR}/L_{\star}$, and the shape of their spectral energy distributions (SEDs). However, SED measurements are dependent upon the properties and evolution of the emitting circumstellar dust, including grain size, composition, and radial distance, as well as albedo, porosity and shape. Therefore, SED modelling results are often poorly constrained, with many physically plausible systems.

\begin{figure}[H]
    \centering
    \includegraphics[width=\textwidth]{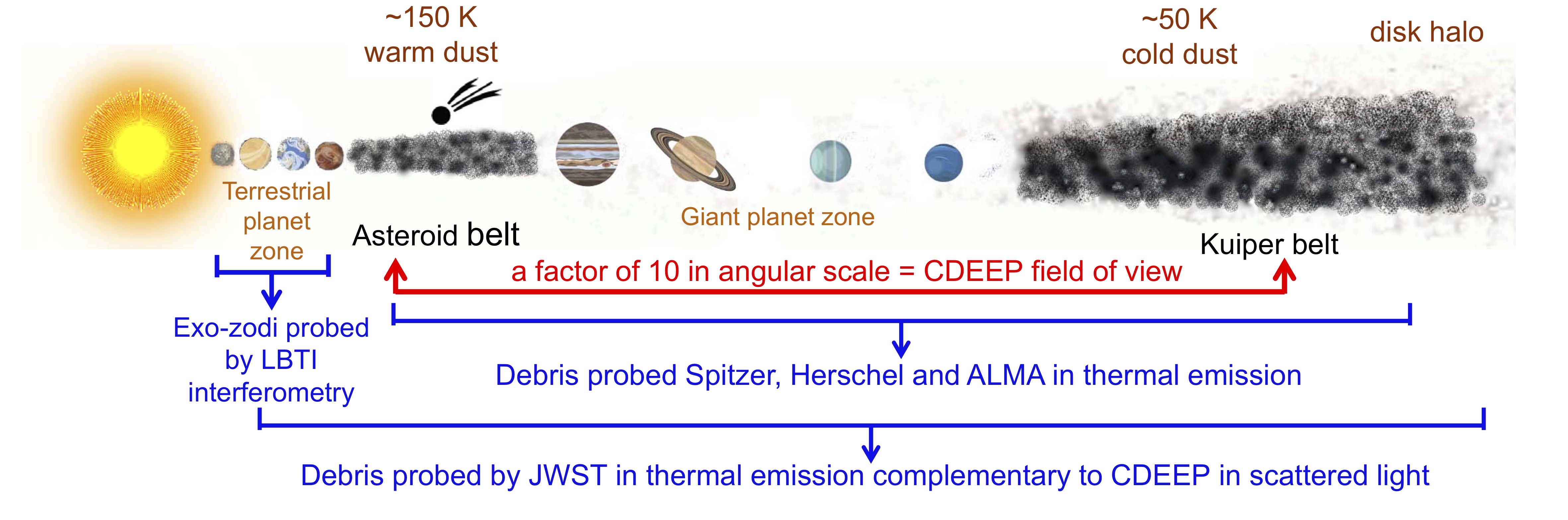}
    \caption{Illustration of the architecture of the solar system where the warm, asteroid-like and the cold, Kuiper-belt-like planetesimal belts are separated by the four giant planets. CDEEP's large field-of-view (FOV), ranging a factor of 10 in angular scale, enables a direct connection between the warm and cold debris in exoplanetary systems. This unique capability probes the debris studied by mostly unresolved IR observations, complementary to the recently completed NASA/Large Binocular Telescope 
    Interferometer (LBTI) exozodi survey\cite{ertel_hosts_2020} and future \emph{James Webb Space Telescope} (JWST) observations.}
    \label{fig:architecture}
\end{figure}

\begin{figure}[b!]
    \centering
    \includegraphics[width=.6\textwidth]{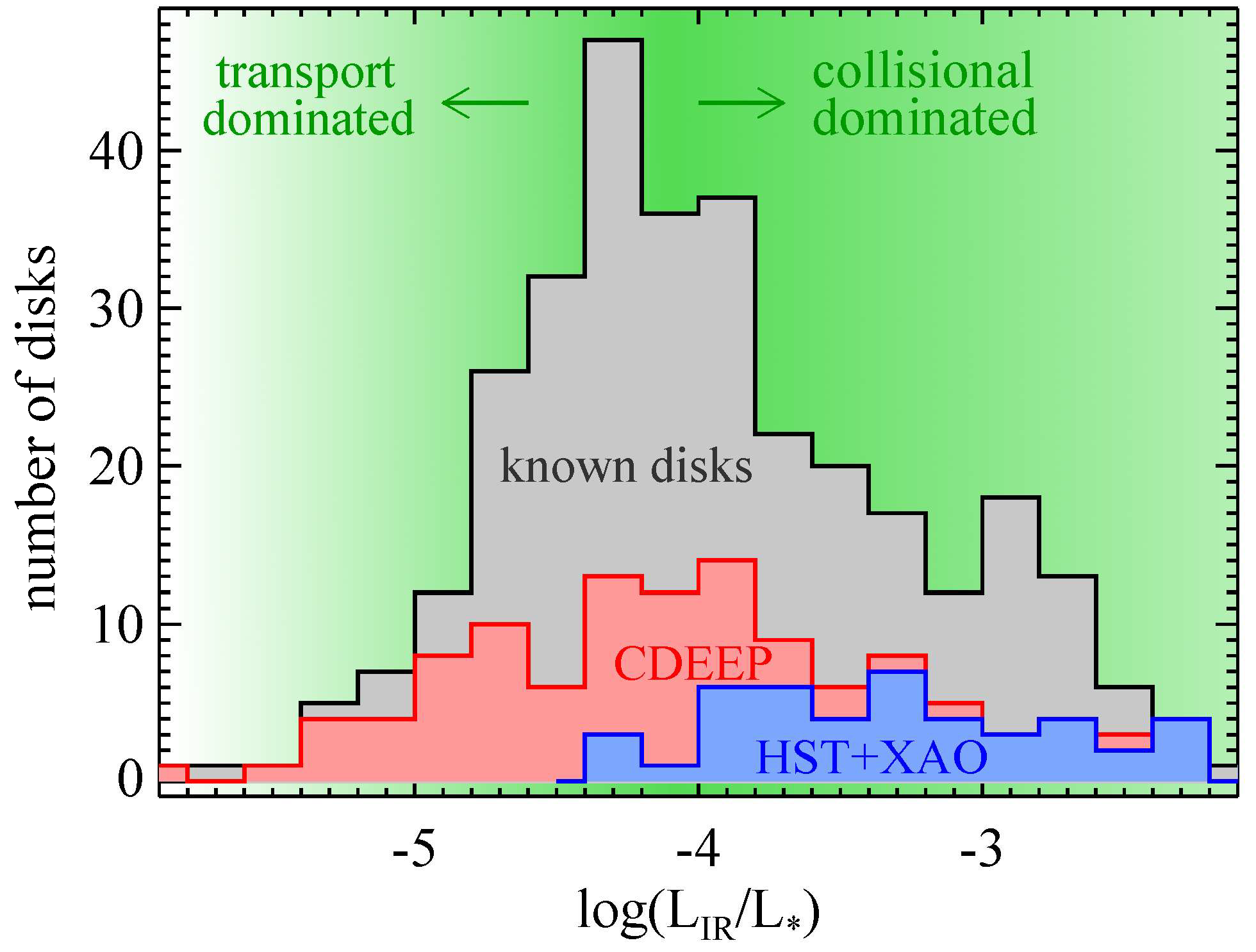}
    \caption{The IR fractional luminosity distribution of known debris disks discovered by \emph{Spitzer} and \emph{Herschel} (gray). The disks which have been resolved in scattered light (blue) belong to the brighter end of the known population, while the CDEEP targets (red) will extend resolved population toward the fainter end, investigating the transition between collision and transport dominated disks.}
    \label{fig:elodie_sample}
\end{figure}

We thus need resolved images to form a cohesive picture of disk morphology and dust properties. Scattered light images are particularly important as they have better spatial resolution than infrared imaging and trace the warmest and coldest components of a disk with a single wavelength. Scattered light imaging of optically-thin disks, coupled with SED measurements of dust emission and resolved thermal images, helps constrain the total mass and location of a circumstellar disk. Observations of both the inner and outer parts of extrasolar systems will probe the connection between cold outer disks and warmer exozodiacal disks (Figure\ \ref{fig:architecture}), search for signs of planets embedded in the disks, and constrain the albedo of the dust -- essential to understanding the diversity of exoplanets detected by current means. Quantifying scattered light brightness in the habitable zone of extrasolar systems is also critical to constraining the background signal which must be overcome by future missions to directly image and characterize exoplanets, sometimes called the exozodi problem \cite{roberge_exozodiacal_2012, kral_exozodiacal_2017,backman_exozodiacal_1998,stark_maximizing_2014,turnbull_search_2012}. However, limited control signals\cite{guyon_limits_2005,males_ground-based_2018} on the ground, and the passive nature of the coronagraphic instrument on HST have prevented the imaging of the vast majority of disks that have been inferred from IR excess. Only a few dozen bright systems have scattered light images, as shown in Figure \ref{fig:elodie_sample}. 

The Coronagraphic Debris and Exoplanet Exploring Pioneer (CDEEP) is a proposed Small-Sat mission concept for high contrast imaging of circumstellar disks in scattered light at visible wavelengths. The CDEEP mission will advance our understanding of exoplanetary system architectures by measuring the brightness, morphology and dust properties of circumstellar debris around nearby stars.

CDEEP will fly fully integrated wavefront sensing/control and coronagraphic technologies in low-Earth orbit, thus also performing important technological pathfinding for upcoming exoplanet imaging missions, for which advanced coronagraphic and wavefront sensing technology have been identified as a critical need. Rigorous testing and characterization of the system in a space-analogous environment and early software development and testing will be essential to the success of the CDEEP mission. To this end, we are developing a vacuum compatible CDEEP prototype at the University of Arizona. In this work we report on the development of the testbed optical design and preliminary simulations of the sensitivity of the system to various errors and the ability of the wavefront sensing and control scheme to correct for them.

\section{CDEEP Mission Overview}

Here we give an overview of CDEEP's proposed science operations scheme and the telescope/instrument payload.

\subsection{Science Operations}

Seventy targets have been identified as candidates for the CDEEP survey. Figure\ \ref{fig:elodie_sample} shows their $L_{IR}/L_\star$ distribution in comparison to all known disks within 300 pc and to those resolved with HST or XAO. The observable CDEEP sample spans the full range of known IR excesses, a wide range of disk sizes, and stars with spectral types evenly divided between early and solar-like stars, at a median age of $\sim$300 Myr (from the literature) and a median distance of $\sim$30 pc. The total number is only limited by target star magnitude and whether the disk size falls within the high-contrast FOV (0.5--6"). The majority of the disks (85\%) will be resolved by at least 10 resolution elements (resel, assuming one resel of 0.32"), sufficient for searching for resonant structures perturbed by unseen planets, and bridging the transition region between transport- and collisionally-dominated regimes \cite{kuchner_collisional_2010}.  


The CDEEP sample includes many nearby iconic debris systems such as Vega, Fomalhaut, $\epsilon$ Eri and $\beta$ Pic where complex disk structures have been intensively studied in thermal IR and some in scattered light. They provide a crucial foundation for interpreting debris disk behavior in terms of the underlying planetary system configuration. Excepting $\beta$ Pic, the inner regions of these benchmark debris systems currently lack scattered-light imaging. Several of the CDEEP targets have NASA/LBTI measurements with exozodi having been detected in the habitable zone for seven of them\cite{ertel_hosts_2020}.
A subset of CDEEP targets are also known to host exoplanets either by RV and/or direct imaging measurements, and this subset is particularly important for studying planet-disk interactions. 

\subsection{Telescope and Instrument Payload}

\begin{figure}[h]
\begin{minipage}[t]{0.4\textwidth}
\includegraphics[width=\textwidth]{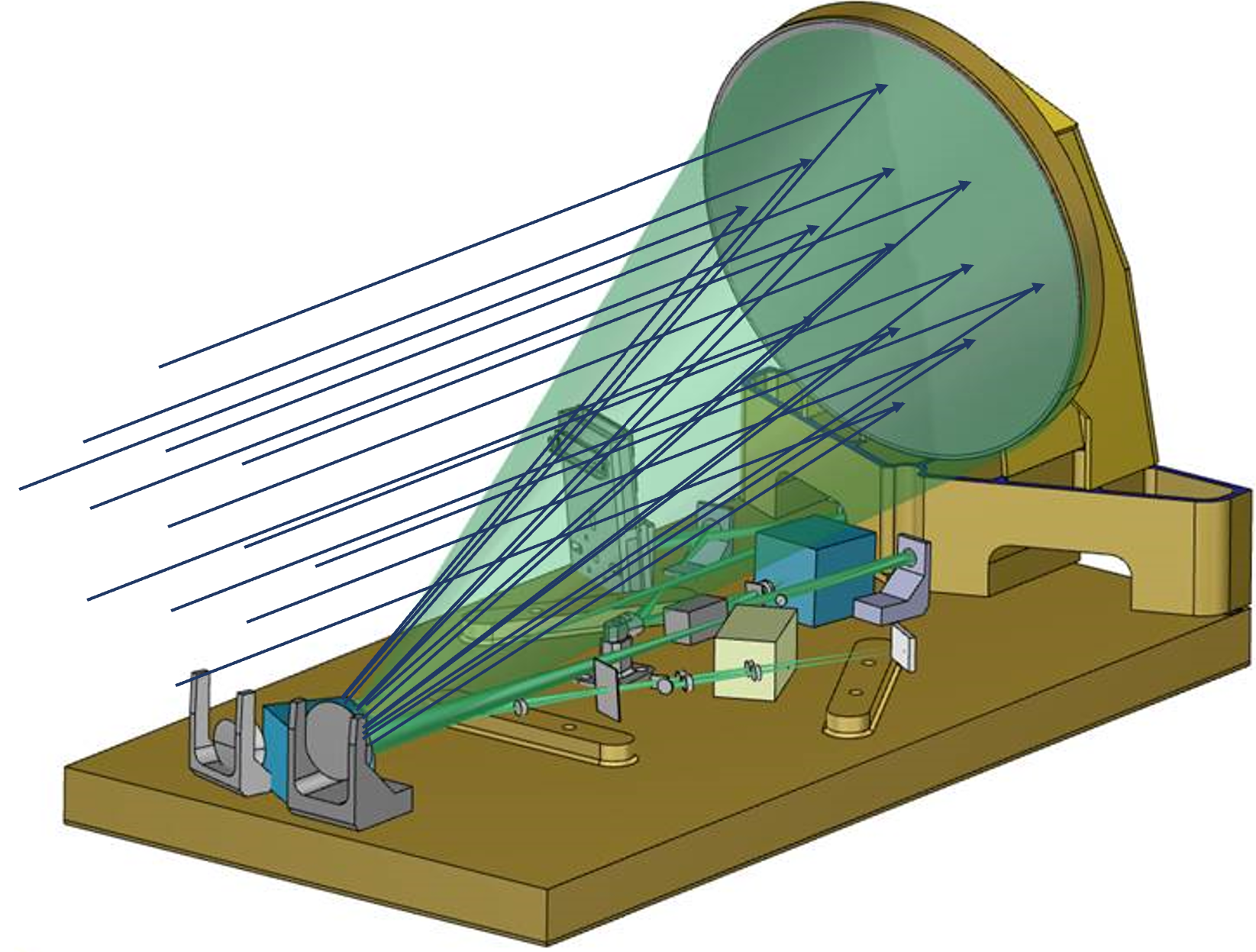}
\caption{Optomechanical rendering of the unobscured SiC telescope and integrated optical bench (rays from the left).}\label{fig:telescope_rendering}
\end{minipage}
\hfill
\begin{minipage}[t]{0.53\textwidth}
\includegraphics[width=\textwidth]{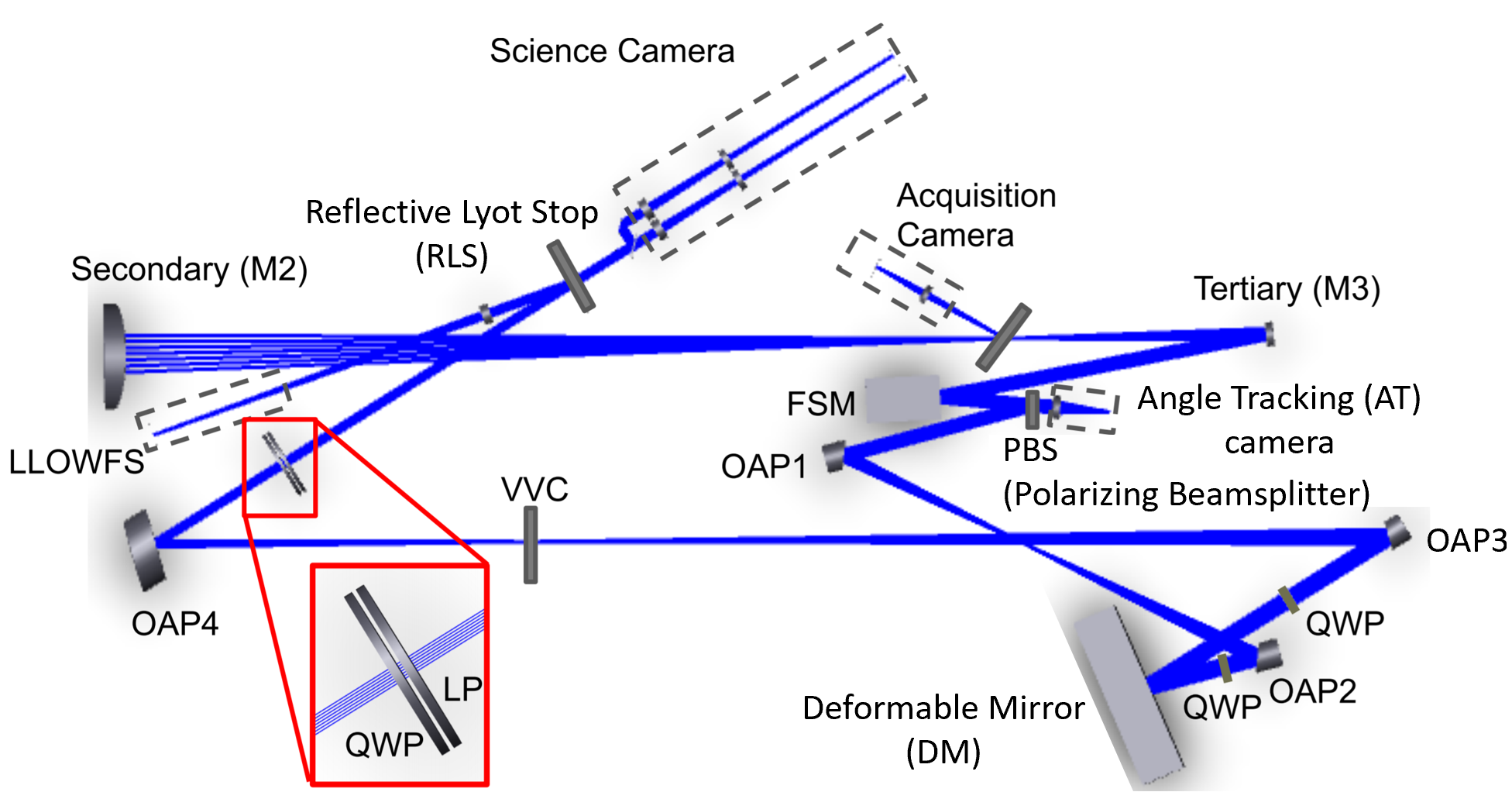}
\caption{ Details of the CDEEP coronagraph optical path (from the M2 on the left) using VVC, LP (Linear Polarizer), and QWP (Quarter Wave Plate).}\label{fig:choi_overall_raytrace}
\end{minipage}
\end{figure}

CDEEP is proposed to be a 34.9 cm off-axis monolithic silicon carbide (SiC) telescope integrated with the coronagraph optical bench, as seen in Figure \ref{fig:telescope_rendering}. CDEEP leverages the heritage of the MEMS Deformable Mirror Demonstration Mission (DeMI)\cite{allan2018deformable,morgan_mems_2019}, the PICTURE series balloon
\cite{mendillo_optical_2017} and sounding rocket missions\cite{mendillo_picture_2012,chakrabarti_planet_2015,douglas_wavefront_2018}, EXCEDE laboratory testing\cite{belikov_excede_2014,sirbu_excede_2015}, and a wealth of theoretical and laboratory work on VVC technology\cite{mawet_optical_2009,mawet_vector_2010,ruane_decadal_2019,serabyn_technology_2019,ruane_2020}. The unobscured off-axis design eliminates diffraction effects from a conventional secondary obscuration and support structure, significantly easing the challenge of high-contrast imaging \cite{mawet_vector_2010-1}. The telescope and coronagraph optical layouts are shown in Figure \ref{fig:choi_overall_raytrace}.
The fast Cassegrain design was chosen to maximize the aperture in the volume, is well matched to the coronagraph size, and can be mounted on a common optical bench, greatly simplifying the metering structuring and minimizing the number of interfaces that must be tracked.
At the telescope focus, a reflective pin-hole aperture provides a large FOV acquisition camera imaging for coarse acquisition by passing only the on-axis beam. 
Fine pointing will be corrected by the Fast Steering Mirror (FSM) with a piezoelectric stage using light rejected by the polarizing beamsplitter between the FSM and OAP1.

The stabilized on-axis beam will reflect off a deformable mirror (DM) within a relay mirror configuration creating a conjugate plane to the system stop M1.
Low-order aberrations are controlled by the DM with feedback from the LLOWFS after the VVC downstream. The ``corrected" complex field is focused to a charge-6 Vector Vortex Coronagraph (VVC)\cite{mawet_optical_2009} using a slow focusing mirror providing a $\sim$f/60 beam at the center of the phase mask. A charge-6 VVC balances relative insensitivity to wavefront errors (Figure \ref{fig:ContrastVSZern}) with small IWA=2.4$\lambda/D$. A linear polarizer followed by a quarter wave plate (QWP) circularizes light into the VVC and a downstream (QWP) and linear polarizer reject non-circularized leakage.
A single, circular polarization state incident improves broadband VVC contrast \cite{murakami_design_2013} and allows polarimetric measurements. 
The phase-modulated beam then propagates to an over-sized  OAP mirror, capturing all the diffracted light field from the on-axis star.
A reflective Lyot stop (RLS) is placed between the over-sized OAP mirror and the science camera. Off-axis debris disk light passes through the RLS while the on-axis starlight is reflected towards the LLOWFS.

\section{Testbed Prototype Design Development}

To accelerate the development of CDEEP's instrumentation and associated software, a prototype testbed has been developed at the University of Arizona. Here we describe the general constraints that have driven its development, as well as the overall contrast budget which is shown in Table \ref{tab:wfe_budget}. 

\subsection{Optical Design Constraints}

Our prototype design was motivated by the goal of demonstrating a coronagraphic testbed that can accurately simulate the behavior of our proposed flight model while minimizing cost. Thus we required:

\begin{enumerate}
    \item All optics should be commercially available off-axis parabolic mirrors (OAPs).
    \item All testbed OAPs must have similar optical powers to their flight model equivalent, and their surface quality should be $<\lambda/20$ RMS.
    \item The total testbed footprint should be no greater than 1x0.75 meters$^2$.
    \item All optics should be mounted with hardware that is suitable for a vacuum chamber (e.g. vacuum-compatible stainless steel) to allow for testing in a space-like environment.
    \item The system (from source to science camera) should maintain diffraction-limited performance.
    \item The system should maximize the beam footprint on the Boston Micromachines Kilo-DM.
    \item The Talbot effect for spatial frequencies of up to 12 cycles/pupil should be constrained. See Section \ref{sec:talbot}.
    \item The system should have room for an out-of-pupil deformable mirror for future testbed development.
    \item The system should include a common path for two sources such that a 4D PhaseCam and a 532nm laser can be used as a source independently without realigning the system.
\end{enumerate}

    \subsubsection{Talbot Effect Control}\label{sec:talbot}
    
    Adopting the approach of N'Diaye et al.\cite{N_Diaye_2013} we consider the ramifications of the Talbot effect on the distances between pupil planes. The Talbot effect is a consequence of 3-beam diffraction on sinusoidal gratings\cite{goodman2005introduction} that results in the self-imaging of the grating at particular distances. For a generally rough optical surface, each spatial frequency across the optic acts like a grating with a different associated Talbot distance. In between the grating and its self-image mixing between amplitude and phase is observed, which is a limiter to high contrast imaging instrument performance\cite{douglas_accelerated_2018-1}. For small pupil sizes (such as those on MEMS DMs) the Talbot effect occurs at relatively short distances, on the order of the specified testbed footprint. Consequently, the distances between optics must be constrained with consideration for the Talbot effect so that minimal mixing between amplitude and phase will occur. We require our testbed distances to be constrained to less than a quarter of a Talbot length ($<<z_{t}/4$) to mitigate the coupling of phase into amplitude. The Talbot distance ($z_{t}$) at a particular frequency ($f_{max}$) is given by: \\
    
        \begin{equation}
            z_{t} = \frac{2*(D/f_{max})^2}{\lambda}
        \end{equation}
        
        Where $D$ is the size of the pupil and $\lambda$ is wavelength. To meet the required contrast of the testbed it is desireable to control 12 cycles per pupil at a wavelength of 532nm.

    \subsection{Contrast Requirements}
    
    The design of the testbed is also motivated by CDEEP's contrast requirements, shown in Table \ref{tab:wfe_budget}. CDEEP's overall raw contrast requirement is $10^{-7}$ per resel, with a post-processing contrast of $10^{-8}$. We have applied substantial conservatism in developing our contrast budget which combines all of the instrument error terms into a raw contrast per resel. The contrast budget is informed by lab tests\cite{prada_high-contrast_2019,serabyn_technology_2019}, sub-orbital-missions\cite{douglas_end--end_2015,mendillo_optical_2017}, and future mission concepts\cite{stapelfeldt2015exo,ruane_decadal_2019,gaudi_habitable_2018}. Compared to these missions, CDEEP has a smaller aperture and thus less sensitivity to pointing and manufacturing errors. The CDEEP primary mirror is of long focal length and thus is less sensitive to polarization aberrations. In Section \ref{sec:results} we describe and present the results of simulations validating those terms which were not identified from literature for the testbed.
    
    \begin{table}[h!]
\centering
\footnotesize
\begin{tabular}{lll}
\toprule
Term & CBE & Reference\\
\midrule
Req. RMS [$1/as^2$] &  $1 \times10^{-6}$ & - \\
Req. RMS$/$resel & $1 \times10^{-7}$ & -\\
\midrule
Polarization       &          $1\times10^{-8}$ & Section \ref{sec:polarization}\\
 DM stability        &          $5\times10^{-9}$ &       \cite{prada_high-contrast_2019} \\
Optical surface     &          $9\times10^{-9}$ & Section \ref{sec:highwfs}\\
Chromatic      error     &          1$\times10^{-9}$ &                 \cite{serabyn_technology_2016} \\
Surface Scatter          &          1e-09 &   \cite{mendillo_optical_2017} \\
\bottomrule\\
\end{tabular}
\caption{Testbed Contrast Budget (2.4$\lambda/D$-10$\lambda/D$ at 540 nm ). 1$\lambda/D\sim0.1$as$^2$. Including on-orbit error terms such as telescope polarization and jitter gives a RMS per resel best estimate for the mission of       $\sim3\times10^{-8}$.}
\label{tab:wfe_budget}
\end{table}

\section{Results}\label{sec:results}
    With the above constraints in mind we present our approach to seeking a cost-effective optical design that meets our performance requirements.

\subsection{Sensitivity Analysis}
 
\begin{figure}[h!]
     \centering
          \includegraphics[width=.8\textwidth]{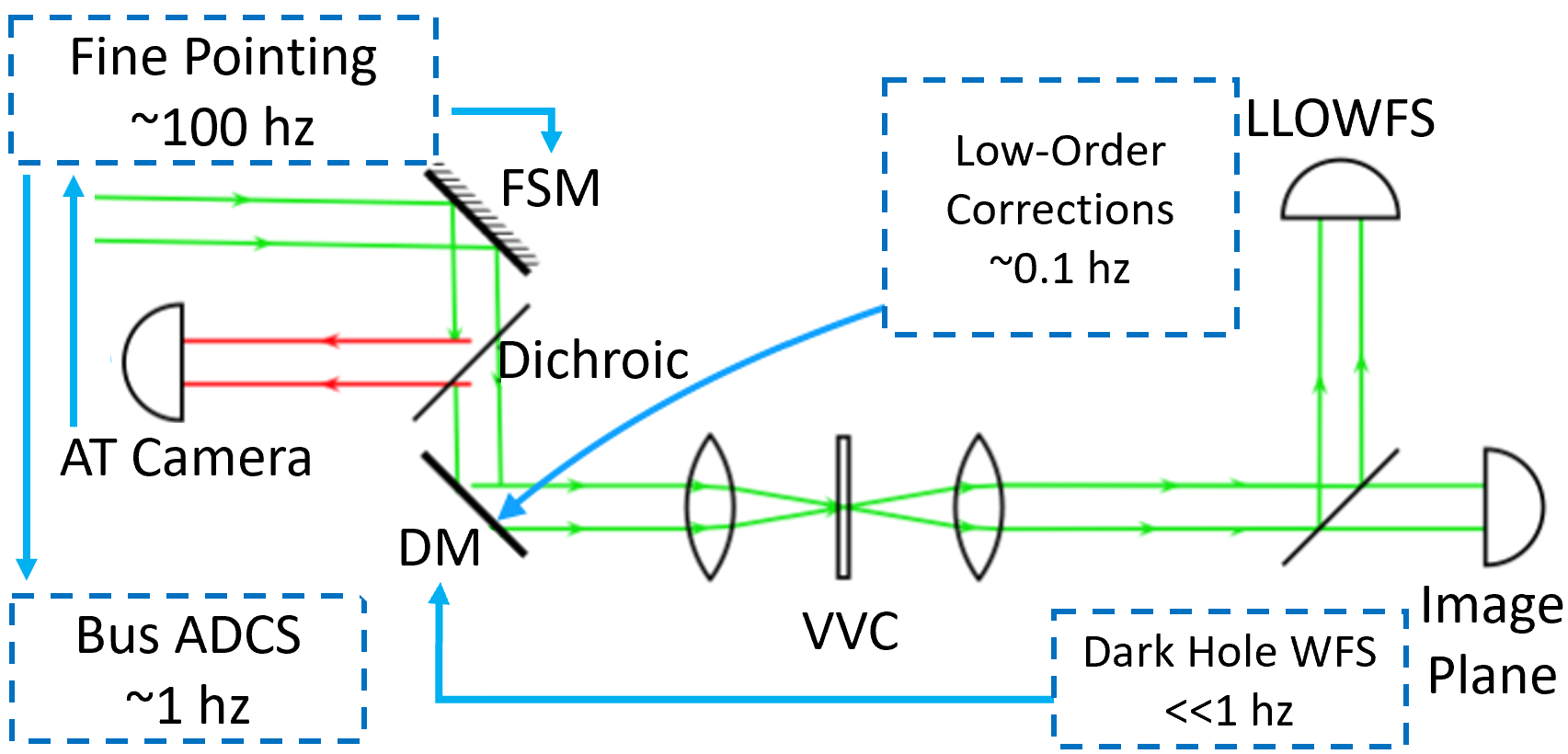}
     \caption{The CDEEP WFSC control system layout includes a high-rate FSM, a high-order MEMS DM fed by a LLOWFS and EFC from the science camera. }
     \label{fig:control_loop}
\end{figure}

To meet the contrast requirements outlined in Table \ref{tab:wfe_budget}, CDEEP's wavefront sensing and control (WFSC) system must be able to correct for a variety of dynamic and static aberrations. CDEEP's overall WFSC layout is shown in Figure \ref{fig:control_loop}. The relevant parts of this system for the prototype testbed will be the fine pointing performed by the FSM, the low-order corrections performed by the LLOWFS, and the high-order corrections performed by the DM and the science camera, which we describe below.

\begin{figure}[ht!]
        \centering
        \includegraphics[width=.9\textwidth]{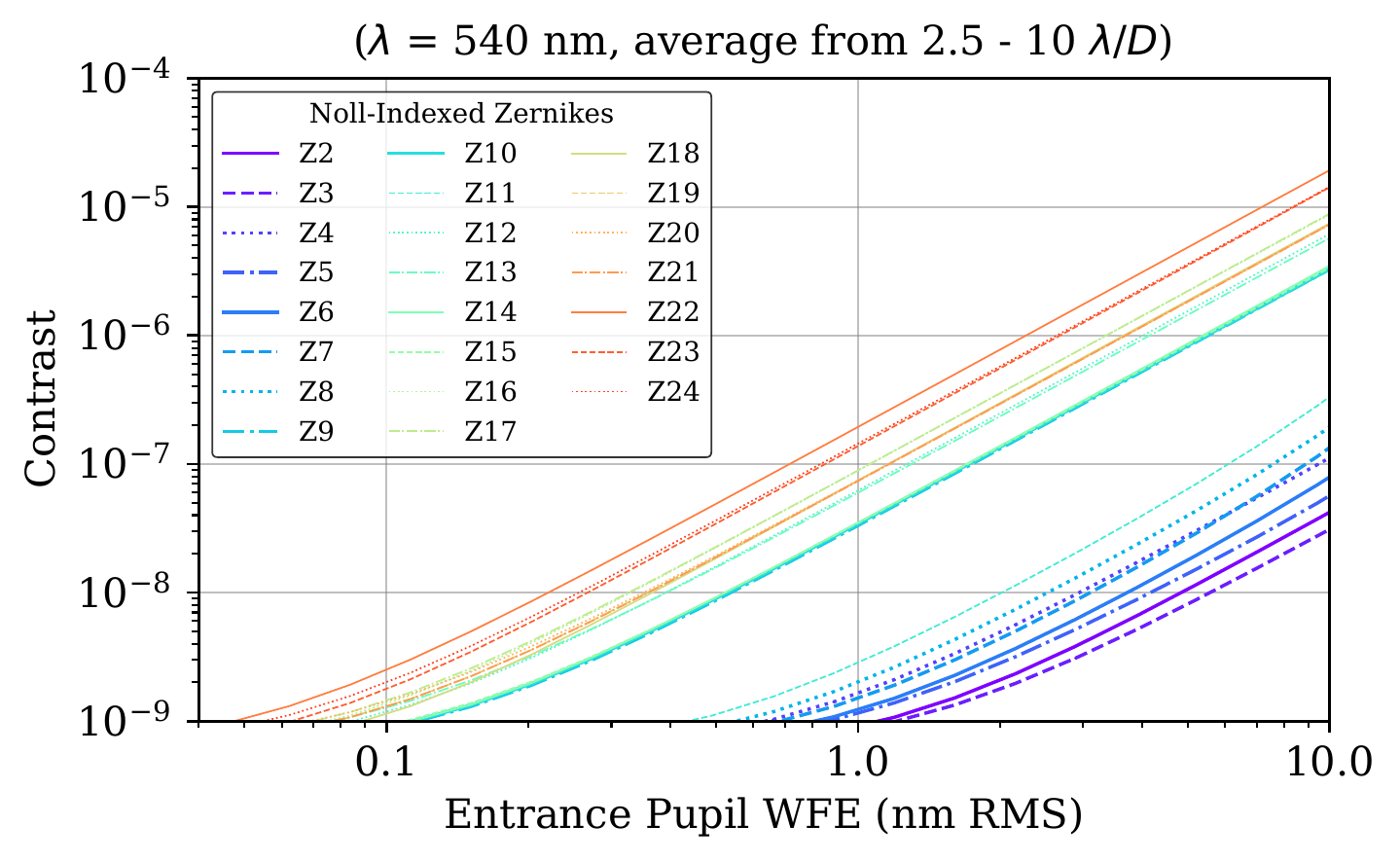}
        \caption{Sensitivity of a charge-6 vector vortex coronagraph to Zernike  aberrations\cite{noll_zernike_1976}.}
        \label{fig:ContrastVSZern}
    \end{figure}

First we conducted an analysis of the general sensitivity of the VVC to aberrations in the incoming wavefront, with the aberrations decomposed using Zernike polynomials. An HCIpy\cite{por_high_2018} model of a vector vortex coronagraph was created, and individual Zernike modes were propagated to the image plane for a range of amplitudes. The average contrast in the dark hole region was retrieved, and this contrast as function of aberration amplitude for 23 Zernike polynomials starting from tip can be seen in Figure \ref{fig:ContrastVSZern}.

        \begin{figure}[ht!]
            \centering
            \includegraphics[width=.9\textwidth]{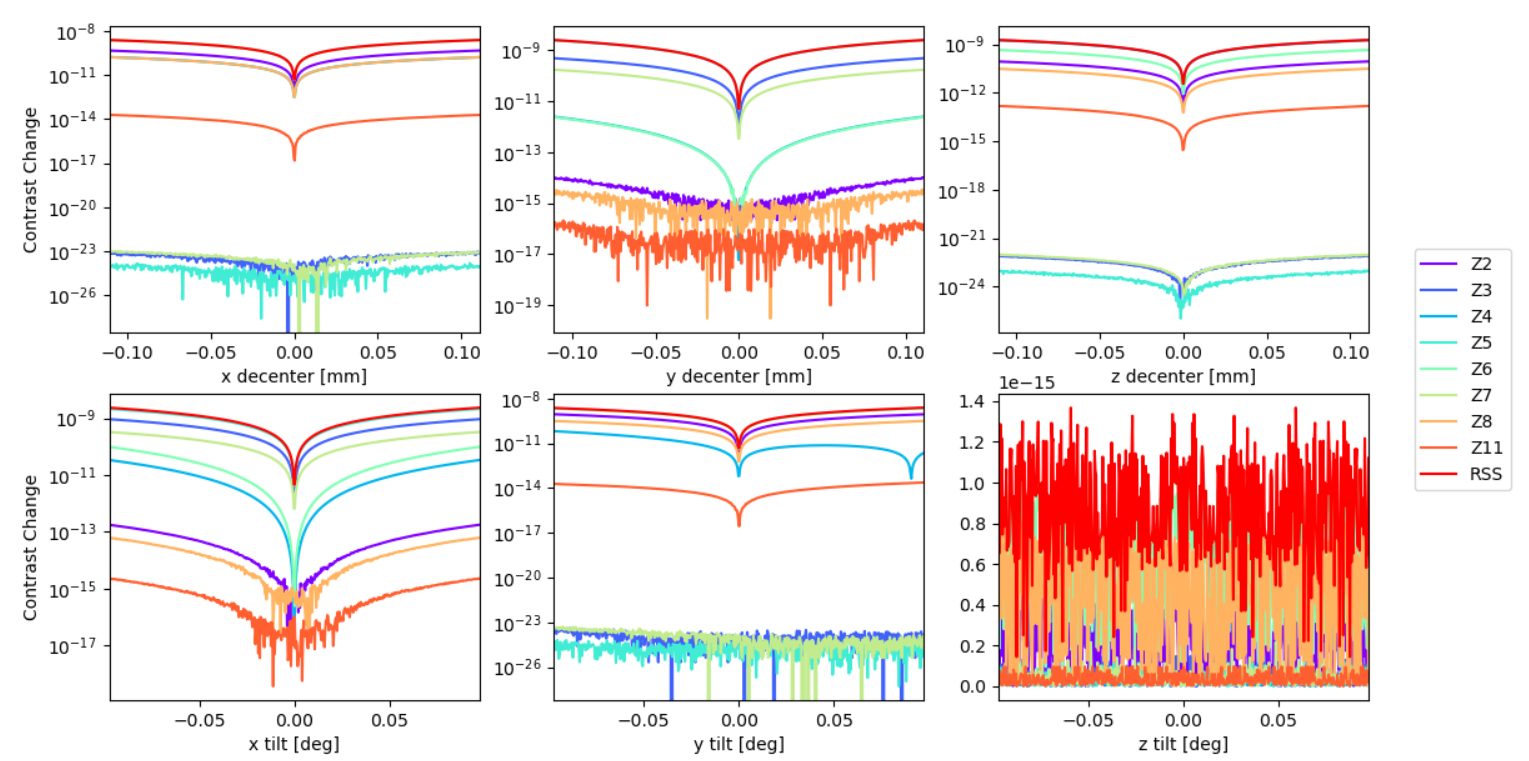}
            \caption{Zernike decomposed sensitivity analysis for the coefficients corresponding to tip, tilt, and the third order monochromatic aberrations. The characteristic feature of this analysis is the gullwing plot shape with a minimum near zero. The range of the x axis of each plot is selected for comparison against the mechanical tolerance of mounting hardware. These plots are used to analyze if a mount will be precise enough to meet the testbed's static contrast requirements ($\sim 10^{-8}$). The RSS (Root Sum Squared error, shown in Red) is plotted with the Zernike coefficients to show the total error contribution.}
            \label{fig:perturb_contrast}
        \end{figure}       

Additionally, sensitivity analyses are useful tools in understanding the geometric wavefront error induced by imperfect assembly and alignment of an optical system. Conventional tools such as Zemax OpticStudio allow sensitivity analysis of performance degradation due to perturbations in six spatial degrees of freedom, but only for typical performance metrics like RMS wavefront error, spot size, etc. 
To inform the mechanical mounting tolerances of the testbed optics for a VVC coronagraph, a numerical model of the coronagraph in HCIpy\cite{por_high_2018} allowed mapping between contrast due to individual Zernike modes induced in the testbed by misalignment errors. 
        
This mapping was used to convert Zernike coefficient values to contrast degradation. To analyze the contrast degradation as a function of tolerance perturbation, a new sensitivity analysis tool was created using OpticStudio's ZOS-API with Python. The six spatial degrees of freedom (x/y/z decenter, tilts) were perturbed individually, and at each step Zernike coefficients were fit to the resultant wavefront. Multiplying the retrieved Zernike coefficients by the data in Figure \ref{fig:ContrastVSZern} yields a map of tolerance perturbation to contrast degradation. Figure \ref{fig:perturb_contrast} shows the sensitivity of M3 with an x-axis chosen for a tip/tilt mount with a resolution of $0.4^{\circ}$ per alignment knob rotation. A quarter turn of the knob corresponds to a $0.1^{\circ}$, which is used as the magnitude of displacement in the X/Y tilt plots (lower left, right, respectively). The RSS error is $<10^{-8}$ at the most extreme tilt, indicating this as a suitable mount for tip/tilt control.

Decomposing the optical element sensitivity into Zernikes also allows consideration for which errors can be corrected using the adaptive optics of the system. For example: static tip (Z2) and tilt (Z3) errors can be compensated for by setting a new baseline tilt on the FSM.

\subsection{Thermal Deformation}\label{sec:thermal}

\begin{figure}[H]
    \centering
    \includegraphics[width=0.85\textwidth]{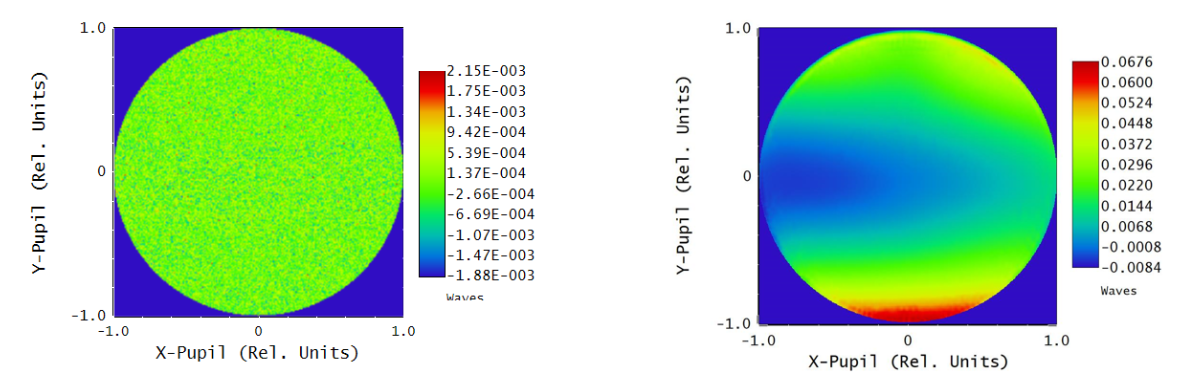}
    \caption{(Left) Wavefront evaluated at the FSM pupil before STOP analysis. (Right) Wavefront evaluated at the FSM pupil after thermal deformation. This worst-case example shows wavefront error without payload thermal control or the LLOWFS correction.}
    \label{fig:stop_star_wavefront}
\end{figure}

High-contrast space coronagraphs require sub-nanometer level stability to suppress starlight over extended integrations. Sufficient knowledge of how the optics deform in response to temperature changes is necessary for modeling the anticipated performance of the testbed, particularly for when it moves into a vacuum environment. To model the surface deformation due to thermal expansion and the resultant wavefront degradation, Zemax OpticStudio was used to import finite element analysis (FEA) data directly onto the mirror surfaces. Figure \ref{fig:stop_star_wavefront} shows the CDEEP wavefront in the flight instrument evaluated at the FSM pupil before and after thermal deformation in a worst case scenario without payload thermal control or correction from the LLOWFS. Simulations also for the flight instrument have shown that the LLOWFS can correct for this and improves contrast by a factor of 100, and we expect even better performance from the testbed due to a more stable thermal environment on the ground.

\subsection{Jitter}\label{sec:jitter}

The telescope pointing must be sufficiently stable that the star stays centered on the coronagraph. Setting pointing leakage to below $1\times10^{-8}$ in the contrast budget requires  $\sim$6 nm RMS for a charge-6 VVC, which is set by the sensitivity to tilt, seen in Figure \ref{fig:ContrastVSZern}. This sets a CDEEP jitter requirement of $\lesssim$ 6 mas, which is easily met by the combination of spacecraft attitude control and the FSM using conventional closed loop control. Further margin is expected using predictive control\cite{males_ground-based_2018} and will be simulated in the lab.

\subsection{High Order WFS}\label{sec:highwfs}

\begin{figure}
    \centering
     \includegraphics[width=.55\linewidth]{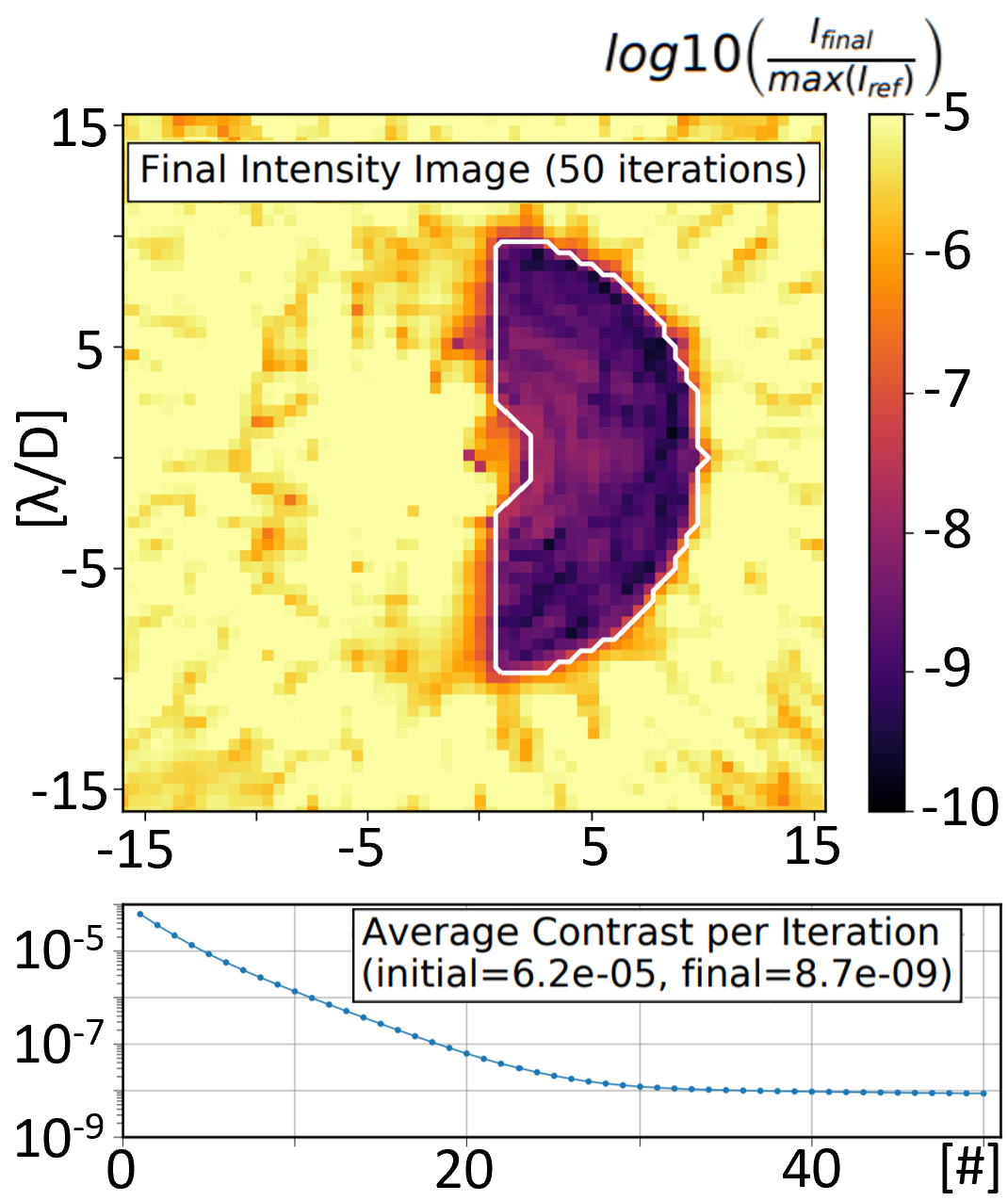}
     \caption{The static wavefront error contrast for CDEEP coronagraph is $10^{-8}$ after $\sim$30 EFC iterations in 10\% broadband light using a Charge-6 scalar vortex coronagraph and $\lambda/10$ PV optics.} 
     \label{fig:vcc_dark_sim}
\end{figure}

High-order wavefront sensing and control for CDEEP will be performed using the science camera via Electric Field Conjugation (EFC)\cite{giveon_electric_2007,groff_methods_2015}. 
A form of image plane wavefront sensing (WFS), EFC uses DM perturbations (``probes") to solve a linear model of the coronagraph system that can be used to iteratively apply DM corrections and suppress residual starlight in the image plane until the desired contrast is reached in the region of interest. Using a single DM to control amplitude and phase will produce a one-sided dark hole as shown in Figure \ref{fig:vcc_dark_sim}.

CDEEP raw contrast is $\sim10^{-2}$ less stringent than that required for ExoEarth  imaging\cite{trauger_laboratory_2007}. Table \ref{tab:wfe_budget} assigns a $10^{-8}$ static contrast floor to uncorrectable manufacturing error after EFC, consistent with past tests and simulations for typical $\lambda/10$ peak-to-valley (PV) or better surfaces\cite{mendillo_optical_2017,jovanovic_high-contrast_2018,llop-sayson_high-contrast_2020}. To validate these assumptions, we performed preliminary modeling of the CDEEP optical train using HCIPY \cite{por_high_2018} and PROPER \cite{krist_proper:_2007}. The optical train was simplified to the DM, a charge-6 scalar vortex coronagraph (in the process of being updated to VVC), the Lyot stop, and the detector. CDEEP baselines two 10\% bandwidth filters centered at 540-nm and 660-nm, simultaneously illuminating the science detector. The secondary blue channel provides simultaneous multi-band imaging to sense disk color and track speckle behavior over time \cite{guyon_spectral_2017} at a smaller IWA. 

We simulated 50 EFC iterations for a 10\% bandpass centered at 540 nm with 120 nm PV random surface errors applied. Broadband EFC was performed as in Belikov (2007)\cite{belikov_demonstration_2007}, where for each iteration we calculated the DM shape optimized for the central wavelength of 540 nm and accepted the contrast degradation in the rest of the bandpass. Thus for each iteration the broadband image was produced by summing together images taken at 10 wavelengths across the bandpass with the DM set to the aforementioned optimized shape. These simulations show that our assumptions are reasonable, and that the required contrast of $10^{-8}$ can be reached in $\sim$30 iterations. We note, however, that these simulations were performed using ``perfect" knowledge of the simulated image plane complex electric field provided by HCIpy, rather than estimates of the field by pairwise probing as will be done for the real system. Thus these numbers thus represent a lower bound and in the real system more iterations will likely be required to reach the required contrast.

\subsubsection{Computational Processing Requirements}

How long would 30 iterations take? EFC is generally a time-consuming process. As briefly described above, it works by making measurements of the complex electric field in the image plane and solving for the DM shape that best negates the electric field, and repeating the process until the desired contrast is reached. This is estimated to take 10 hours for the Roman telescope\cite{bailey_lessons_2018}. However, the actuator number for CDEEP is $<1/4$ that of Roman, proportionally decreasing the control problem dimensions. Additionally, the CDEEP processor, currently specified as the quad core, 1.2GHz Xiphos Q8S, is expected to reduce the convergence time to of order twenty minutes, as shown below. CDEEP will also require fewer iterations to reach the lower required contrast and machine learning training on the ground will accelerate convergence\cite{sun_efficient_2020}. 


We can calculate the estimated execution time for various procedures involved in an EFC iteration\cite{Pogorelyuk}. Table \ref{tab:timeperop} reports the estimated time required for the Xiphos Q8S board to perform a single operation in nanoseconds. This assumed that only the 4 ARM A53 cores in the SOC would be used to run the algorithm and there is no OS or abstraction layer between the calculations and the CPU. The ``best case" is wildly unrealistic and only looks at the raw computing performance of the CPU with no consideration for memory access or the difficulty of parallelizing a task over multiple cores. The ``worst case" reflects a deliberately un-optimized system (use of cache turned off forcing the CPU to access main memory frequently, etc).

Table \ref{tab:totalcomptime} reports the estimated execution time in seconds for various procedures involved in an EFC iteration, based on estimated number of operations. These calculations are performed for a 180 degree dark hole imaged at 3 different wavelengths and sampled at 0.5$\lambda$/D from 3-15 $\lambda$/D, which equates to approximately 1500 pixels, with a 1024 actuator DM. In brief, these procedures are:

\begin{enumerate}
    \item Pair-probing step: Estimating (``probing") of the image plane electric field.
    \item Classical EFC step: Calculating the DM actuator voltages that best negate the estimated image plane electric field. The number of operations required can be reduced by an order of magnitude by point 3:
    \item Using a pre-computed gain matrix and a fixed regularization parameter for the required inversion of the complex response matrix that relates DM shape to image plane intensity, but this requires point 4:
    \item Computing the gain matrix. 
\end{enumerate}

Thus, the total time required to perform EFC for N iterations would be either (1 + 2) * N, if not using a precomputed gain matrix, or 4 + ((2 + 3) * N). For 50 iterations, in the worst case scenario, with calculating and using the gain matrix, this would again take of order 20 minutes.

\begin{table}[H]
\centering
\begin{tabular}{|c|c|}
\hline
\textbf{Case} & \textbf{ns per Operation} \\ \hline
Worst Case & 18.83333 \\ \hline
Single Core Best Guess & 4.22333 \\ \hline
Multi Core Best Guess & 2.63958 \\ \hline
Best Case & 0.09057 \\ \hline
\end{tabular}\\
\caption{Estimated time per operation for the Xiphos Q8S board in nanoseconds for several different cases.}
\label{tab:timeperop}
\end{table}

\begin{table}[H]
\centering
\begin{tabular}{|c|c|c|c|c|c|}
\hline
\textbf{Step} & \textbf{\begin{tabular}[c]{@{}c@{}}Number of \\ Operations\end{tabular}} & \textbf{Worst Case} & \textbf{\begin{tabular}[c]{@{}c@{}}Single Core\\  Best Guess\end{tabular}} & \textbf{\begin{tabular}[c]{@{}c@{}}Multi-core\\ Best Guess\end{tabular}} & \textbf{Best Case} \\ \hline
Classical EFC step & 5.20E+08 & 9.79E+00 & 2.20E+00 & 1.37E+00 & 4.71E-02 \\ \hline
Pair-probing step & 8.5E+05 & 1.60E-02 & 3.59E-03 & 2.24E-03 & 7.70E-05 \\ \hline
\begin{tabular}[c]{@{}c@{}}Fixed regularization parameter\\ with precomputed gain matrix\end{tabular} & 1.8E+07 & 3.39E-01 & 7.60E-02 & 4.75E-02 & 1.63E-03 \\ \hline
Pre-computed EFC gains & 1.90E+10 & 3.58E+02 & 8.02E+01 & 5.02E+01 & 1.72E+00 \\ \hline
\end{tabular}\\
\caption{Total estimated computation time in seconds for each of the 4 steps outlined in the text for each of the above cases.}
\label{tab:totalcomptime}
\end{table}

\subsection{Polarization}\label{sec:polarization}

Mixing between polarization states leads to wavefront errors which can not be corrected by the DM\cite{mendillo_polarization_2019}.
A polarization ray trace of optics upstream of the VVC was employed to quantify these errors. Polarization ray tracing (PRT) is a technique for calculating the polarization matrices for ray paths through optical systems. Polaris-M was built from the ground up to calculate polarization effects in optical systems. It is based on a 3x3 polarization ray-tracing calculus.
Diffraction image formation of polarization aberration beams is then handled by vector extensions to diffraction theory\cite{chipman_lam_young_2009}.

\begin{figure}[H]
     \centering
     \begin{subfigure}[t]{0.45\textwidth}
         \centering
         \includegraphics[width=\textwidth]{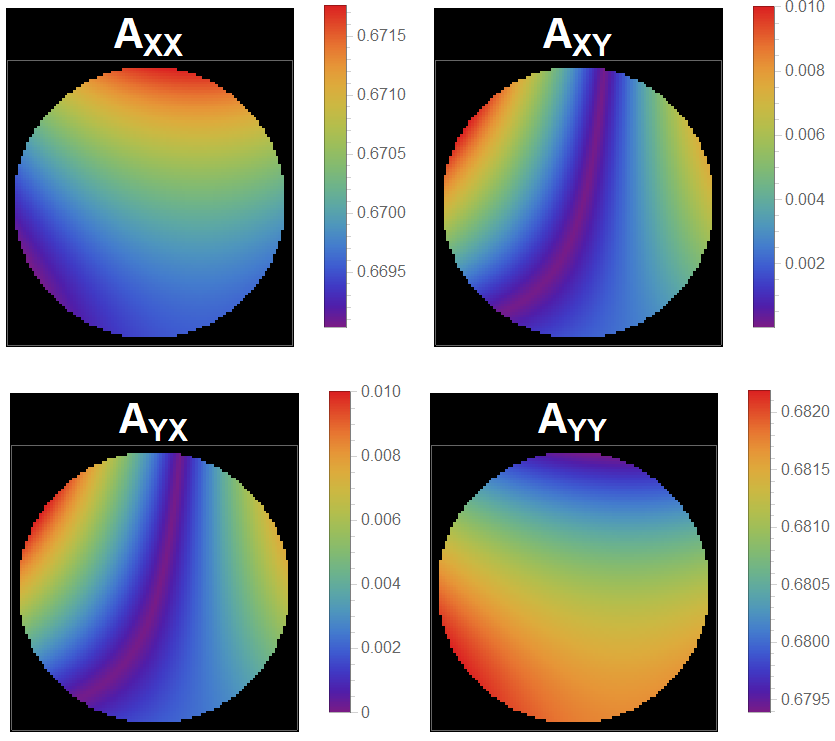}
         \caption{Jones Pupil magnitude entering the VVC before inserting polarizing elements. Magnitude shows roughly 70\% of light passes through the system into the VVC. The off-diagonals values show negligible light leakage, with maximum transmission of 1\% x-polarized light into y and vice-versa.}
         \label{fig:JP_MagNoPol.PNG}
     \end{subfigure}
     \hfill
     \begin{subfigure}[t]{0.45\textwidth}
         \centering
         \includegraphics[width=\textwidth]{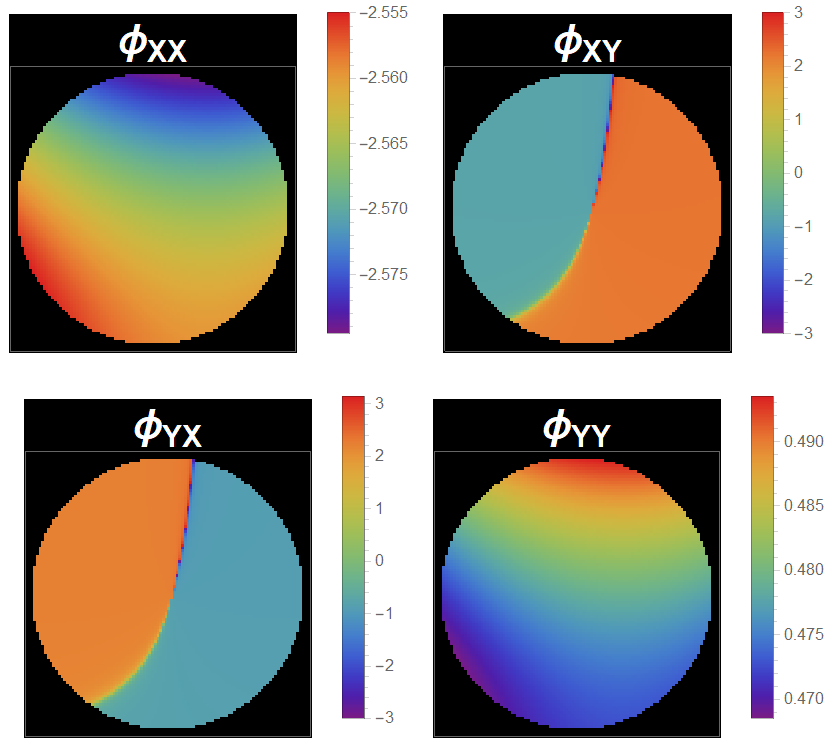}
         \caption{Jones Pupil phase entering the VVC before inserting polarizing elements. On-diagonals show phase difference of ~$\pi$, while the off diagonals are close to identical.}
         \label{fig:JP_ArgNoPol.PNG}
     \end{subfigure}
\end{figure}

Optics upstream of the vector vortex coronagraph were modeled to observe the Jones Pupil incident on the focal plane vortex mask. The resulting Jones Pupil of the first half of the system is shown in Figure \ref{fig:JP_MagNoPol.PNG} and \ref{fig:JP_ArgNoPol.PNG}. 
The contrast contribution, calculated by fitting Zernikes to the off-diagonal terms of the Jones matrix, of the telescope and relay optics is $<10^{-8}$; thus, the testbed including the relay optics, lacking the telescope, is expected to have negligible polarization error.

\subsection{Optics Specification and Final Design}

 After a period of downselect, a vendor was identified with available master parabolic mirrors which would allow OAPs with focal lengths similar enough to the flight model and with the required surface quality and Talbot lengths to be produced. The specifications of the testbed OAPs are reported in Table \ref{tab:oapspecs} and their fractional Talbot lengths are reported in Table \ref{tab:talbot}. To further decrease polarization and alignment error, the off-axis distances (OADs) of our OAPs were slightly reduced and modeled with vacuum-compatible mounts in SolidWorks at each design iteration to ensure the mounting hardware would not vignette the coronagraph beam.
        
        \begin{table}[H]
            \centering
            \begin{tabular}{|c|c|c|}
                \hline
                Optic & Radius of Curvature & Off-Axis Distance Change From Flight Model to Testbed  \\
                \hline
                M3 & -293.6mm & 0mm \\ \hline
                OAP1 & -254mm & 35mm \\ \hline
                OAP2 & -346mm & 42mm \\ \hline  
                OAP3 & -914.4mm & 154mm \\
                \hline
            \end{tabular}
            \caption{Mirror prescriptions for the Off-Axis Parabolas upstream of the VVC in the testbed.}
            \label{tab:oapspecs}
        \end{table}
        
    \begin{table}[H]
        \begin{center}
        \begin{tabular}{|c|c|c|c|}
        
            \hline
            Distance (Optic 1 - Optic 2) & Value [mm] & $z$/$z_{T}$ \\ 
            \hline
             M3-FSM & 145  & 0.16\\ \hline
             FSM-OAP1 & 126 & 0.14 \\ \hline
             OAP1-OAP2 & 312 & 0.18 \\ \hline 
             OAP2-DM & 182 & 0.10 \\
            \hline
        \end{tabular}
        \caption{The distances between optics upstream of the deformable mirror expressed in units of mm and fractional talbot distance $z/z_t$.}
        \label{tab:talbot}
        \end{center}
    \end{table}

    
    
\begin{figure}[h!]
    \centering
    \includegraphics[width=\textwidth]{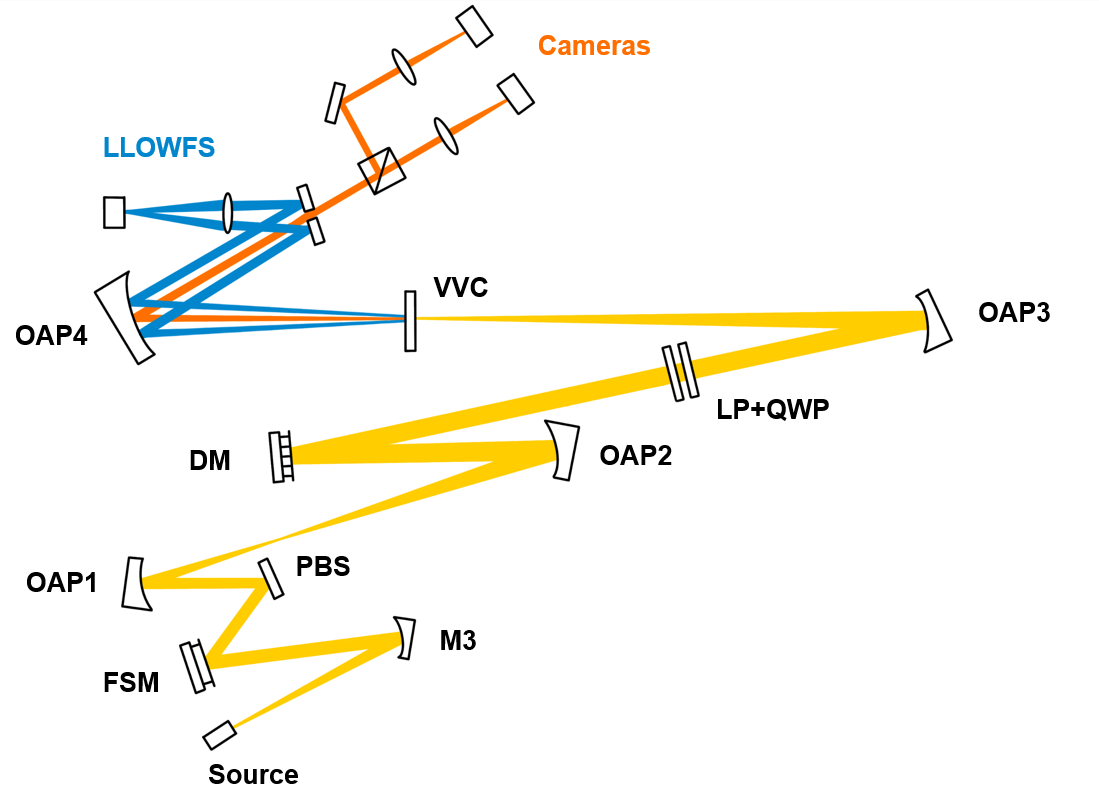}
    \caption{The optical design for the CDEEP prototype testbed. The optical train begins at the source on the bottom left. The blue beams downstream of the VVC represent the starlight that is rejected by the reflective Lyot stop and used by the LLOWFS, while the orange represents the debris disk light that continues to the science cameras.}
    \label{fig:testbed}
\end{figure}

\begin{figure}
    \centering
    \includegraphics[width=\textwidth]{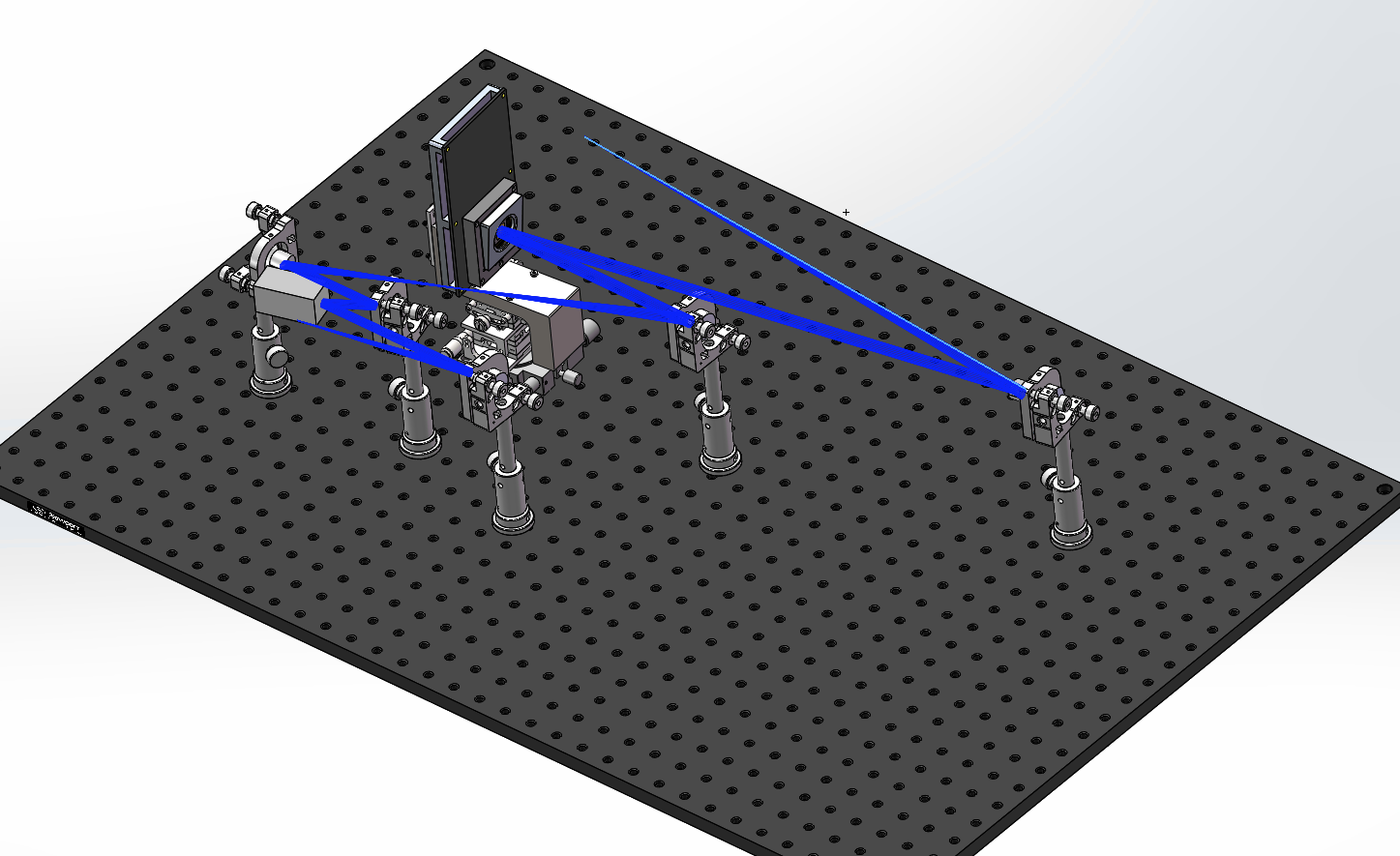}
    \caption{3D rendering of the testbed optomechanical system as seen in schematic in Figure \ref{fig:testbed}.}
    \label{fig:testbed_render}
\end{figure}

 The final optical design of the prototype testbed can be seen in schematic in Figure \ref{fig:testbed} and as a 3D rendering in Figure \ref{fig:testbed_render}.
 
\section{Conclusions and Future Work}

Here we have presented an overview of the proposed CDEEP missions and the science it will address. We have presented the design of the under-construction vacuum prototype testbed that will be used to accelerate software development and allow testing of a close to flight equivalent system in a space-like environment. We present simulations validating the ability of the proposed testbed design and wavefront sensing and control scheme to reach several requirements laid out in the overall CDEEP contrast budget, including contributions from polarization error, optical surface error, thermal deformation, jitter, and tolerance to alignment error.

Ongoing work includes selection of remaining components for the testbed, including the FSM and vacuum-compatible optics mounts, and some refinement of the optical layout to ensure optimum performance. Simulations are also being performed to compare the performance of the proposed design of CDEEP to a larger on-axis design. Future work will include further characterization of the processing requirements of EFC, as well as simulations of its performance when using realistic pairwise probing vs. perfect simulated knowledge of the image plane electric field. 
\acknowledgments 

Many thanks to the coauthors of the CDEEP proposal for their hard work on said proposal in a very difficult time, and to those who gave feedback on this work. 
Portions of this work were supported by the Arizona Board of Regents Technology Research Initiative Fund (TRIF). 
Thanks also to Hop Bailey and the Arizona Space Institute for their support of the mission concept.
An allocation of computer time from the UA Research Computing High Performance Computing (HPC) at the University of Arizona is gratefully acknowledged.
Thanks to Zemax for the access to their OpticStudio STOP analysis feature. 
Special thanks to everyone at the Space Telescope Science Institute Russell B. Makidon Optics Laboratory for their advice.

\bibliography{report,exoplanets,section2} 
\bibliographystyle{spiebib} 

\end{document}